# Model-Free Reinforcement Learning for Automated Fluid Administration in Critical Care

Elham Estiri
College of Aeronautics and Engineering
Kent State University
Kent, OH, USA
eestiri@kent.edu

Hossein Mirinejad
College of Aeronautics and Engineering
Kent State University
Kent, OH, USA
hmiri@kent.edu

*Abstract*— Fluid administration, also called fluid resuscitation, is a medical treatment to restore the lost blood volume and optimize cardiac functions in critical care scenarios such as burn, hemorrhage, and septic shock. Automated fluid administration systems (AFAS), a potential means to improve the treatment, employ computational control algorithms to automatically adjust optimal fluid infusion dosages by targeting physiological variables (e.g., blood volume or blood pressure). Most of the existing AFAS control algorithms are model-based approaches, and their performance is highly dependent on the model accuracy, making them less desirable in real-world care of critically ill patients due to complexity and variability of modeling patients' physiological states. This work presents a novel model-free reinforcement learning (RL) approach for the control of fluid infusion dosages in AFAS systems. The proposed RL agent learns to adjust the blood volume to a desired value by choosing the optimal infusion dosages using a Q-learning algorithm. The RL agent learns the optimal actions by interacting with the environment (without having the knowledge of system dynamics). The proposed methodology (i) overcomes the need for a precise mathematical model in AFAS systems and (ii) provides a robust performance in rejecting clinical noises and reaching desired hemodynamic states, as will be shown by simulation results.

## I. Introduction

Intravenous fluid administration is a crucial treatment to restore the blood volume (BV) and stabilize critically ill patients in hypovolemic scenarios. Successful fluid management depends on the type and dosage of infusion. A small volume of fluid may be insufficient to meet the desired outcomes, whereas an over-aggressive infusion regimen may result in severe medical complications [1]. Automated fluid administration systems (AFAS) are decision-making algorithms that automatically adjust the fluid infusion dosages based on the value of hemodynamic endpoints such as BV or mean arterial pressure (MAP) [2], [3]. AFAS systems have the potential to reduce the incidence of human errors in clinical settings, lower the risk of under- and over-dosing in fluid resuscitation, and act as the supportive care to save lives during public health emergencies [4].

Model-based approaches for AFAS systems have been studied in the last decade [2], [4], [5], [6], and [7]. A comparison study between the performance of a fuzzy logic and a decision table AFAS controller was performed for different bleeding scenarios in [2]. In [5], a closed-loop learning intravenous fluid resuscitation algorithm was developed to optimize the cardiac output as the hemodynamic endpoint. In [6], a proportional-integral (PI) control-oriented model was designed to replicate the change of BV in hemorrhagic scenarios. In [7], a model-based adaptive control algorithm was developed to regulate the MAP using a hemodynamic model relating blood pressure to the fluid gain and loss. The main drawback of model-based control approaches is that the performance of the controller depends on the accuracy of the model. Recent studies have indicated that the existing model-based approaches have difficulties in finding accurate dose-response models for AFAS systems mainly due to complexity and variability of modeling patients' physiological states and the lack of a robust identification tool to deal with uncertainties such as clinical noises [4], [8]. A model-free approach based on machine learning techniques may address this issue.

Reinforcement learning (RL) is a machine learning approach for the control of complex uncertain, dynamical systems [9]. An RL agent can learn the optimal actions without having the knowledge of system dynamics by interacting with the environment. Model-free RL has shown promising results in the areas such as robotics [10], transportation [11], air traffic management [12] and recently healthcare domain, including anemia management [13]-[16], insulin therapy [17], closed-loop anesthesia [18], and MAP control [19].

Of particular interest, the application of RL in medical autonomy was promising: In [13], an RL-based anemia management algorithm was proposed to maintain the hemoglobin concentration within the target range. In [17], some of the key challenges related to the automated insulin therapy was addressed using the RL. In [18], an RL-based algorithm was successfully designed to keep the bispectral index (BIS) and MAP in the desired range for ICU patients. In [19], a closed-loop MAP control system was designed by the RL for critical care patients.

In this work, we present a model-free RL control approach for regulating fluid infusion dosages in bleeding scenarios. To the best of Authors' knowledge, this is the first attempt at applying RL to AFAS systems. We designed a Q-learning algorithm that proposes an optimal drug delivery schedule for fluid administration. The proposed approach (i) overcomes the need for a precise mathematical model in AFAS systems and (ii) provides a robust performance in rejecting clinical noises and reaching the desired hemodynamic state, as will be shown in Section III.

The remainder of the paper is organized as follows: Section II describes the model used as the virtual patient generator in this study, as well as the proposed model-free RL methodology; Simulation results are shown in Section III; Discussions are presented in Section IV; And finally, conclusions are drawn in Section V.

This material is based upon work supported by the National Science Foundation under Grant No. 2138929.

## II. MATERIALS AND METHOD

### A. Hemodynamic Model

A lumped-parameter model of BV, characterizing the patient's response to fluid infusion, was developed in [6]. This control-oriented model represents the fluid shift between the intravascular and interstitial compartments, which can be formulated as

$$\ddot{\tilde{V}}_B(t) + k\dot{\tilde{V}}_B(t) = \frac{1}{V_{B0}}[\dot{u} - \dot{v}] + \frac{k}{V_{B0}(1+\alpha)}[u - v] \quad (1)$$

where $\alpha$ is target volume ratio between the intravascular and extravascular volumetric changes in response to fluid gain and loss, $k$ is the feedback gain or the speed of fluid shift between the two compartments, $V_{B0}$ is the initial BV, $\tilde{V}_B(t)$ is the normalized value of change of BV (i.e., $\tilde{V}_B(t) = (V_B(t) - V_{B0})/V_{B0}$), $u$ is the fluid infusion rate, and $v$ represents the rate of fluid loss due to hemorrhage. Also, $\dot{\tilde{V}}_B$ and $\ddot{\tilde{V}}_B(t)$ represent the first and second derivative of $\tilde{V}_B$ with respect to time (for more information about the model, see [6]).

The proposed RL approach is model-free. We used the model of (1) to represent patients' response to fluid changes. In other words, in the absence of a real patient (environment) to interact with the RL agent, the hemodynamic model was used to simulate virtual patients and generate the input/output data.

### B. Model-Free Reinforcement Learning Control

RL, a type of machine learning method, is concerned with learning an optimal behavior by an agent to obtain maximum reward when exploring a dynamic environment. A general RL framework is shown in Fig. 1. Given the state of the environment, an RL agent chooses an action and receives a reward (either positive or negative). This choice of action determines the next state of the environment and affects the next action taken by the agent, iteratively. Over time, the RL agent learns to choose an action that maximizes the sum of the rewards in the long term.

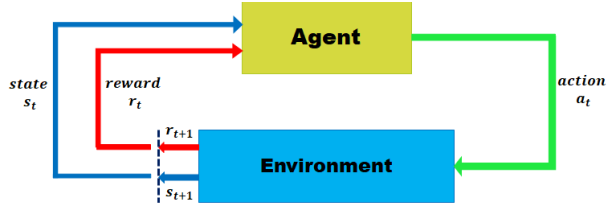

Fig. 1. The reinforcement learning (RL) framework

Finding a series of infusion dosages to track a desired hemodynamic endpoint can be considered as a sequential, goal-oriented, decision-making problem represented by a finite Markov decision process (MDP). A finite MDP can be defined as a 4-tuple $(S, A, P, R)$, where $s_i \in S$ is a set of environment states, $a_i \in A$ is a set of actions taken by an agent, $P$ is the state transition probability matrix with $P_a(s_i, s_{i+1})$ representing the probability of transitioning from $s_i$ to $s_{i+1}$ under an action $a_i$, and $R_a(s_i, s_{i+1})$ is the associated reward function received by action $a$ during the transition from $s_i$ to $s_{i+1}$ (action desirability).

RL methods are naturally suitable for solving MDP problems. In a Q-learning algorithm [9], an RL agent learns how to behave optimally using the action-value function, also called Q-value, $Q: S \times A \rightarrow R$. Q-value function is defined as the expected sum of rewards from action $a$ in state $s_i$ and can be updated during transition $(s, a, r, \acute{s})$ as [9]:

$$Q(s, a) \leftarrow (1 - \gamma)Q(s, a) + \gamma \times [r + \mu \max_{a_{i+1} \in A} Q(\acute{s}, \acute{a})] \quad (2)$$

where $\gamma \in [0,1)$ is the learning rate and $\mu \in (0,1)$ is the discount factor for future rewards. Smaller values of $\mu$ highlight the importance of immediate rewards, whereas the larger values of $\mu$ signify the future rewards.

In the initial phase of learning process, the RL agent uses $\epsilon$-greedy policy to choose actions while exploring the environment. In an $\epsilon$-greedy policy, the agent performs random actions with the probability of $\epsilon$ and chooses an action with the highest Q-value with the probability of $(1 - \epsilon)$.

The state value at time step $k$ is a function of $e(k)$ which is defined as follows:

$$e(kT) = |BV(kT) - BV_{target}| \quad (3)$$

where $BV(kT)$ is the BV at step k, T is the sampling time, and $BV_{target}$ is the desired (target) BV. The reward function is also a function of $e(kT)$ as follows:

$$r_{k+1} = \begin{cases} \frac{e(kT) - e((k+1)T)}{e(k)}, & e((k+1)T) < e(kT) \\ 0, & e((k+1)T) \geq e(kT) \end{cases} \quad (4)$$

According to (4), the RL agent receives a positive reward when the error is decreasing and receives a 0 reward when the error is increasing. After training the agent, it develops an action selection policy that relies on the learnt action-value Q-function as defined by

$$a(k) = \underset{a \in A}{\mathrm{argmax}}\, Q(s(k), a). \quad (5)$$

## III. SIMULATION RESULTS

The model-free RL controller was designed in Python for various fluid administration cases. Simulation results for a virtual patient are demonstrated here. This scenario was incorporated from [1] where a moderate hypovolemia was applied to volunteer human subjects by withdrawal of 900 mL blood prior to fluid administration. Baseline and target BV were set to 3,940 mL and 5,000 mL, respectively. Infusion dosages of ringer's acetate were limited to be between 0 and 25 mL/kg, a maximum dosage derived from [2]. The simulation was run for 100 minutes, and 30,000 episodes were recorded.

An episode represents a series of state-action pairs starting from an arbitrary initial value to the final desired state. The state mapping table is designed based on the value of the error shown in Table I. During the training, it is desired for the RL agent to meet all states in one episode. Once the agent is trained and an optimal Q-value function is obtained, the agent stops exploring the environment and chooses actions with the best former performances using the optimal Q-value function. The RL action set was defined corresponding to the different infusion dosages, i.e., A = [0, 5, 10, 15, 20, 25]. Simulations were conducted by setting $\epsilon = 0.5$ (for $\epsilon$-greedy policy), $\gamma = 0.69$, and $\Delta\tilde{V}_B(t) = 0$. The discount factor $\mu = 0.2$ was assigned initially and halved every 1000 episodes, indicating that the agent tended to choose actions with immediate reward over time. The performance of the RL controller was compared against a proportional-integral-derivative (PID) controller [8] in two simulated cases: without and with observational errors. MATLAB PID Tuner app was used to tune PID gains.

TABLE I. THE STATE MAPPING TABLE

| BV error < 0 | | BV error > 0 | |
|---|---|---|---|
| State number | $e(kT)$ (L) | State number | $e(kT)$ (L) |
| 1 | [0,0.01) | 11 | [0,0.01) |
| 2 | [0.01,0.03) | 12 | [0.01,0.03) |
| 3 | [0.03,0.06) | 13 | [0.03,0.06) |
| 4 | [0.06,0.150) | 14 | [0.06,0.150) |
| 5 | [0.0150,0.400) | 15 | [0.0150,0.400) |
| 6 | [0.0400,0.700) | 16 | [0.0400,0.700) |
| 7 | [0.0700,1.100) | 17 | [0.0700,1.100) |
| 8 | [1.100,1.500) | 18 | [1.100,∞) |
| 9 | [1.500,2.00) | | |
| 10 | [2.000,∞) | | |

*A. BV Measurement without Observational Error*

Fig. 2 demonstrates the achieved BV and the recommended fluid infusion dosages from the RL and PID without observational errors. Achieved BV of the RL algorithm increased from the initial state to the target level with a smooth transition in less than 60 minutes. The infusion started with reasonably safe, low dosages and increased to the maximum permissible rate to take the BV to the desired level in the shortest amount of time. After reaching the BV target, the infusion decreased significantly to the minimum level and the BV remained in its desired range. After reaching the BV target, the infusion rate of RL method fluctuated between 0 and 20 ml/kg every few minutes to keep the BV at the desired level. Compared to the RL-based algorithm, the PID controller spent more time at the maximum infusion level. Also, the steady-state BV from the PID was higher than the desired level, which might be an indicator of overdosing in clinical settings.

*B. BV Measurement with Observational Error*

A white noise with the maximum amplitude of ±250 mL was applied to the output (BV) to simulate the clinical noises caused by different measurement methods, e.g., gravimetric, bag calibration, or dye dilution technique [3]. As demonstrated in Fig. 3a, the proposed controller was capable of achieving the desired BV, despite fluctuations caused by the white noise. In contrast, the BV level from the PID reached a value higher than the BV target (overdosing), as shown in Fig. 3b. Also, comparing Figs. 3b and 2b indicates that the infusion dosages recommended by the RL were the same as those suggested without measurement errors verifying the robustness of the proposed algorithm against the measurement noise. However, the infusion regimen of the PID was negatively affected by the clinical noise, as shown by the fluctuations in Fig. 3b.

*C. Performance Assessment*

To further investigate the efficacy of the proposed algorithm in closed-loop control of BV, the median performance error (MDPE), median absolute performance error (MDAPE), and root mean square error (RMSE) were used as the performance metrics. MDPE, the observed control bias, is defined as

$$MDPE = median(PE(i)), \ i = 1, ..., N \quad (6)$$

where $PE$ is the performance error computed as

$$PE = \frac{BV(t) - BV_{target}}{BV_{target}} \times 100, i = 1, ..., N \quad (7)$$

where $N$ represent the number of BV measurements during the simulation. Also, $MDAPE$ and $RMSE$ are calculated as

$$MDPAE = median(|PE(i)|), i = 1, ..., N \quad (8)$$

$$RMSE = \sqrt{\frac{\sum_{i=1}^{N}(BV(t) - BV_{target})^2}{N}}. \quad (9)$$

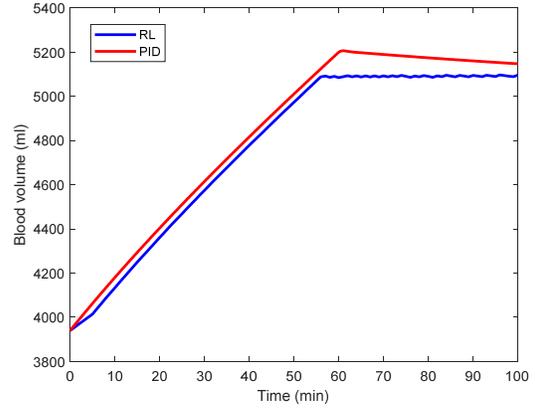

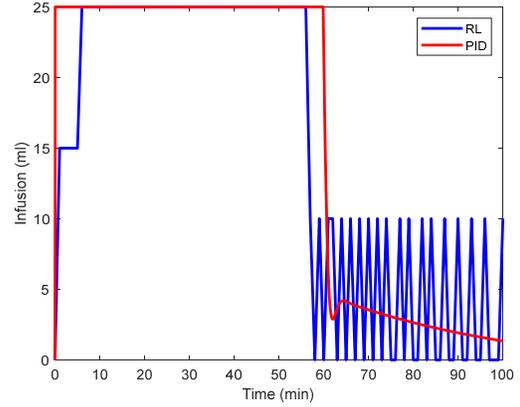

(b)

Fig. 2. (a) Achieved blood volume levels and (b) Fluid dose adjustments from the RL and PID controllers without BV measurement error

Table II shows the performance metrics computed for both RL and PID during the simulation. The results of Table II clearly indicate the superior performance of RL compared to the PID, in terms of all three performance metrics.

TABLE II. PERFORMANCE ASSESSMENT OF THE RL AND PID

| Performance metrics | RL | PID |
|---|---|---|
| $MDPE$ (%) | 0.37 | 1.75 |
| $MDAPE$ (%) | 0.37 | 1.87 |
| $RMSE$ (L) | 0.50 | 0.53 |

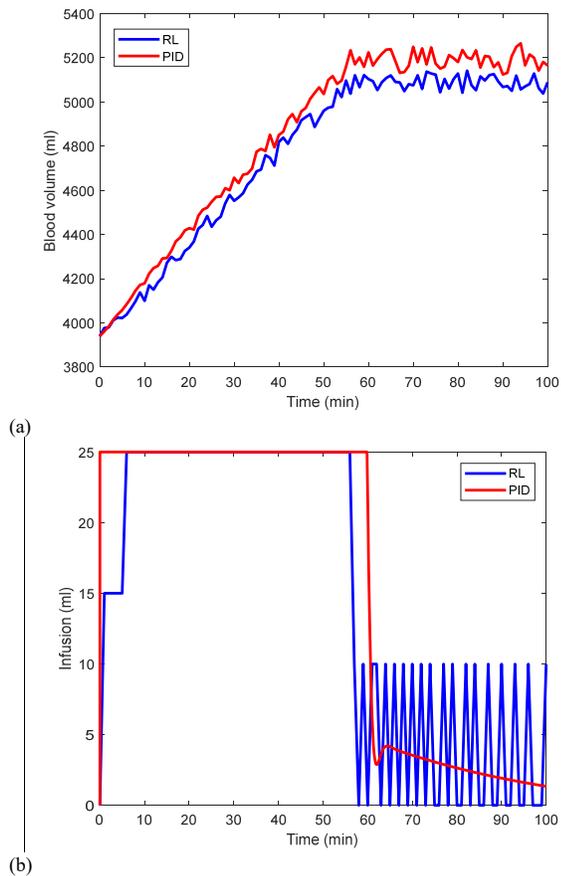

Fig. 3. (a) Achieved blood volume levels and (b) Fluid dose adjustments from the RL and PID controllers with BV measurement error

## IV. Discussion

We proposed a model-free RL algorithm to control the fluid infusion dosages in critical care scenarios. The proposed RL agent develops an optimal action selection policy using the Q-learning algorithm without having an explicit knowledge about system dynamics. Simulation results showed a smooth transition of BV from the initial to the desired state in the presence and absence of observational error. Comparison studies between the RL and PID demonstrated the higher performance of the RL in terms of *MDPE*, *MDAPE*, and *RMSE* performance metrics.

This study considered BV as the design endpoint. The performance of the model-free RL for other hemodynamic endpoints such as MAP and cardiac output can be examined in the future to further assess the proposed algorithm. Leveraging multiple hemodynamic endpoints in multiple medication infusions (e.g., fluid and vasopressors) are a common practice in critical care. Extending the proposed approach to multiple medication infusion scenarios will enable observing the effect of drug interactions in critical care patients. While simulation results are promising, further investigations are needed to assess the robustness of the algorithm against parameter uncertainties and clinical disturbances. In addition, the RL-based controller can be optimized and fine-tuned with a hardware-in-the-loop fluid administration test bed [4], [20] for feasibility assessment and performance testing. Further, the potential of Model-free RL methods to address inter-patient variability in dose-response relationships can be thoroughly examined in future studies.

## V. Conclusions

A model-free RL-based algorithm was designed for the automated control of fluid administration in hemorrhagic scenarios. The simulation results demonstrated the superior performance of the proposed controller compared to the PID in presence and absence of the observational errors. Extending the proposed method to other hemodynamic endpoints and evaluating its performance against other fluid administration controllers will be studied in the near future.